\documentclass[twocolumn,aps,prl] {revtex4}
\usepackage{graphicx,subfigure,latexsym,wasysym,amssymb}

\begin{document}

\title{Effective Viscosity of Confined Hydrocarbons}

\author{I.M. Sivebaek$^{1,2,3}$, V.N. Samoilov$^{1,4}$ and B.N.J. Persson$^1$}
\affiliation{$^1$IFF, FZ-J\"ulich, 52425 J\"ulich, Germany}
\affiliation{$^2$Novo Nordisk A/S, Research and Development, DK-3400 Hillerod, Denmark}
\affiliation{$^3$Mech. Eng. Dept., Technical University of Denmark, DK-2800 Lyngby, Denmark}
\affiliation{$^4$Physics Faculty, Moscow State University, 117234 Moscow, Russia}

\begin{abstract}

We present Molecular Dynamics (MD) friction calculations for confined hydrocarbon films with
molecular lengths from 20 to 1400 carbon atoms.
We find that the logarithm of the effective viscosity $\eta_{\rm eff}$ for nanometer-thin
films depends linearly on the logarithm of the shear rate: ${\rm log}\eta_{\rm eff} = C- n{\rm log}\dot \gamma$,
where $n$ varies from 1 (solid-like friction) at very
low temperatures to 0 (Newtonian liquid) at very high temperatures, following an inverse
sigmoidal curve. Only the shortest chain molecules melt, whereas the longer ones only show a softening
in the studied temperature interval $0<T<900 \ {\rm K}$.
The results are important for the frictional properties of very thin (nanometer) films
and to estimate their thermal durability.

\end{abstract}
\maketitle


The frictional and rheological properties of thin confined fluid films are of great importance
in a number of engineering and scientific applications, e.g., in the context of
lubrication. Thus, recent studies \cite{PS} have shown that
for sliding contacts lubricated by organic or silicon oil bulk shear thinning can have a great influence on
the friction at intermediate sliding velocities (mixed lubrication).
When wetting fluids are confined between solid walls at nanometer separation they often
acquire solid-like properties and an increasing squeezing force is necessary in order to reduce the
film thickness. If the solid walls are smooth the fluid molecules arrange in layers parallel to the
solid walls and the squeeze-out occurs in a quantized way by removing one monolayer after another
with increasing pressure \cite{Klein_1998, Mugele1, Bureau_2010}. Sometimes the last one or two monolayers are so strongly bound
that they cannot be removed by squeezing alone.
Fluid films confined at the nanometer length scale exhibit viscosity enhancement and nonlinear flow properties
characteristic of sheared supercooled liquids approaching its glass transition \cite{Demetriou, Furukawa}.

Experiments for a large variety of fluids (including hydrocarbon
fluids and silicon oil) \cite{Yamada_2002,Bureau_2010} have shown that
the logarithm of the effective viscosity $\eta_{\rm eff}$ for nanometer-thin
films (typically 3 or 4 monolayers and contact pressures of the order of a few
MPa) depends linearly on the logarithm (here and elsewhere with 10 as the basis) of the shear rate: ${\rm log}\eta_{\rm eff} = C- n{\rm log}\dot \gamma$.
If $\eta_{\rm eff}$ and $\dot \gamma$ are measured in SI units, for a
large variety of fluids (at room temperature) $C\approx 5$
and $n\approx 0.9$. This linear relation has also been established in other experiments \cite {Granick_1991, Luengo_1997}
and in computer simulations \cite{Thompson_1992, Rob_Baj_2000}.

We have performed a very extensive set of Molecular Dynamics (MD) simulations to probe the
frictional properties of thin layers of confined hydrocarbon
molecules (with molecular lengths from 20 to 1400 carbon atoms). Some of these results
may also be relevant for polymer-on-polymer systems as recent studies \cite{Yew} have shown
that in this case the shear deformations are localized to a band of material
about $2.5 \ {\rm nm}$ thick.
Our results for $n$ and $C$ agree with the experimental observation at room temperature, but show that
when the temperature increases $n$ varies from 1 (solid-like friction) at very
low temperatures to 0 (Newtonian liquid) at very high temperatures.

Our model is similar to those described in Refs.
\cite{SivebaekEPJE2008,SivebaekLangmuir2010}.
We consider a block and a substrate with
atomically flat surfaces separated by a polymer slab consisting of
hydrocarbons with molecular lengths 20, 100 and 1400 carbon atoms.
The solid walls are treated as single layers of ``atoms'' bound
to rigid flat surfaces by springs corresponding to the
long-range elastic properties of $50 \ {\rm \AA}$
thick solid slabs similar to our earlier papers (see, e.g. Refs. \cite{pervzn2002, SSP2003,S2004}).
The simulation box in $x$-$y$ dimension is equal to $124.8 \ {\rm \AA} \times 124.8 \ {\rm
\AA}$. In the following, periodic boundary conditions are assumed in the $xy$ plane.

Initially about half of the molecules are adsorbed on the block surface and
half on the substrate surface.
Two solids with adsorbed polymer slabs were put into
contact and when the temperature was equal to the thermostat temperature everywhere
we started to move the upper block surface.
The temperature was varied from 0 K
to 900 K to study the effect of temperature (and also melting) on the shear stress.
In our simulations, the polymer films are very thin ($\sim 3 \ {\rm nm}$), and the solid walls
are connected to a thermostat at a short distance from the polymer slab. Under these
circumstances we find that frictional heating effects are not important, and the effective temperature
in the polymer film is always close to the thermostat temperature.

Linear alkanes C$_n$H$_{2n+2}$ (with $n$ = 20, 100 and 1400) were used as
``lubricant'' in the present calculations.
The CH$_{2}$/CH$_{3}$ beads are treated
in the united atom representation \cite{jorgensen1984x1,dysthe2000x1}.
The Lennard-Jones potential
was used to model the interaction between beads of different chains
$$U(r)=4 \epsilon_0 \left [ \left ({r_0\over r}\right )^{12}-
\left ({r_0 \over r}\right )^{6} \right] \eqno(1)$$
and the same potential with modified parameters
$(\epsilon_1,r_1)$
was used for the interaction of each bead with the substrate and block atoms.

The parameters were $\epsilon_0 = 5.12 \ {\rm meV}$ for both the
interior and the end beads, and $r_0 = 3.905 \ {\rm \AA}$.
For the interactions within the C$_n$H$_{2n+2}$ molecules we used the standard optimized potentials for liquid simulations
model \cite{jorgensen1984x1,dysthe2000x1}, including flexible bonds, bond
bending and torsion interaction, which results in bulk properties in good agreement
with experimental data far below the boiling point \cite{Martin1998}.
Atomic mass 14 (for interior CH$_2$ beads) and 15 (for the CH$_3$ end groups) were used.
Within a C$_n$H$_{2n+2}$ chain we assume nearest
neighbor C atoms are connected via springs with the spring constant $k$,
which was chosen equal to $10 \ {\rm N/m}$. Note that this value is one
order of magnitude smaller then the optimized 450 N/m \cite{jorgensen1984x1},
and was chosen such to facilitate a reasonable time step of 1 fs.
We used an angle bending interaction of the form
$E(\cos \theta)/k_B=(1/2) k_{\text{bend}}(\cos \theta-\cos \theta_0)^2$
with $k_{\text{bend}}=62543$K and $\theta_0=2.0001$ rad.
For the dihedral interaction we used the functional form in term
of a cosine Fourier series $E(\phi)/k_B=\sum_{i=0}^3 c_i \cos^i (\phi)$
with parameters $c_0=1009.99$K, $c_1=2018.95$K, $c_2=136.37$K, $c_3=-3165.30$K.
Internal beads of separation greater than 3 units are treated similarly as beads from different chains.
The number of molecules was equal to 1000, 200 and 14
for the C$_{20}$H$_{42}$, C$_{100}$H$_{202}$ and C$_{1400}$H$_{2802}$
systems respectively.
The hydrocarbon films at room temperature consisted of 6 to 8
monolayers of molecules between the solid surfaces.
The (nominal) squeezing pressure $p_0$ was usually $10 \ {\rm MPa}$.

We have chosen the polymer-wall atom bond to be so strong that no
slip occurs at these interfaces. This is the case with $r_1 = 2.92 \ {\rm \AA}$,
$\epsilon_1 = 160 \ {\rm meV}$.
The lattice spacings of the block and of the substrate are $a=b=2.6~\mbox{\AA}$.

\begin{figure}
  \includegraphics[width=0.45\textwidth, angle=0]{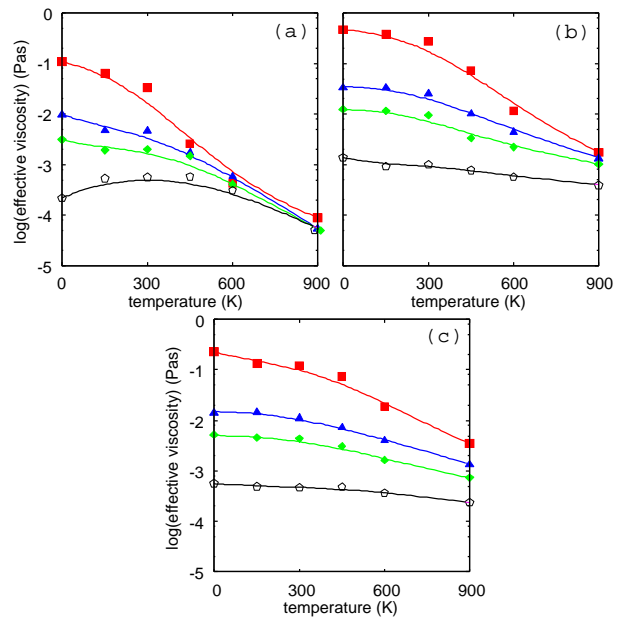}
  \caption{ \label{visc_vel_temp}
(Color online) The logarithm of the effective viscosity as a function of the temperature for (a) ${\rm C}_{20}{\rm H}_{42}$,
(b) ${\rm C}_{100}{\rm H}_{202}$ and (c) ${\rm C}_{1400}{\rm H}_{2802}$
system at four sliding velocities: 0.3 m/s ($\blacksquare$), 3 m/s ($\blacktriangle$), 10 m/s ($\blacklozenge$)
and 100 m/s (\pentagon).
}
\end{figure}

If $v$ is the sliding velocity and $d$ the film thickness, we define the shear rate $\dot \gamma = v/d$ and
the effective viscosity $\eta_{\rm eff} = \sigma_{\rm f} /\dot \gamma$, where $\sigma_{\rm f}$ is the
frictional shear stress.
Fig. \ref{visc_vel_temp} shows the
logarithm of the effective viscosity as a function of the temperature \cite{Temp} for the (a) ${\rm C}_{20}{\rm H}_{42}$,
(b) ${\rm C}_{100}{\rm H}_{202}$ and (c) ${\rm C}_{1400}{\rm H}_{2802}$
system, at four sliding velocities (from top to bottom) $0.3$, $3$, $10$
and $100 \ {\rm m/s}$. Increasing the velocity results in a reduced effective viscosity, i.e., the thin
films exhibit shear thinning.

\begin{figure}
  \includegraphics[width=0.3\textwidth, angle=0]{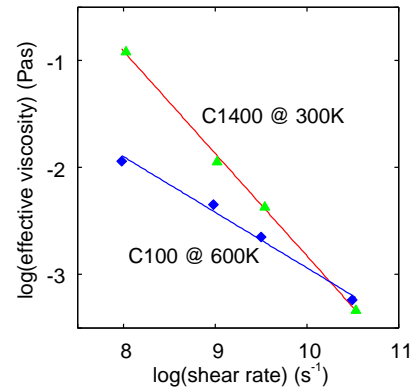}
  \caption{ \label{fit}
(Color online) The data points for the logarithm of the effective viscosity as a function of the logarithm of the shear rate at four sliding velocities:
0.3, 3, 10 and $100 \ {\rm m/s}$. The lines represent the linear fits to these data points. The systems are ${\rm C}_{100}{\rm H}_{202}$
at 600 K and ${\rm C}_{1400}{\rm H}_{2802}$ at 300 K. The values of the slope and the intercept of the ${\rm C}_{100}{\rm H}_{202}$ are
$-0.520 \pm 0.038$ and $2.260 \pm 0.351$ and for the ${\rm C}_{1400}{\rm H}_{2802}$ system $-0.958 \pm 0.024$ and $6.753 \pm 0.228$.
}
\end{figure}

\begin{figure}
  \includegraphics[width=0.3\textwidth, angle=0]{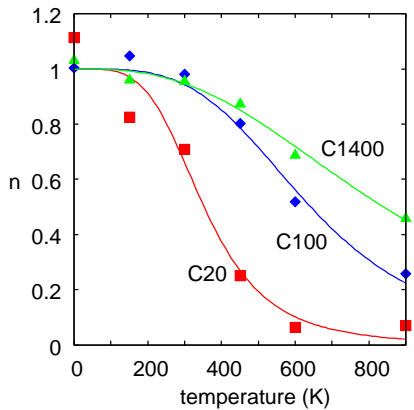}
  \caption{ \label{n_T}
(Color online) Values of the parameter $n$ in equation (2) as a function of the temperature for
the systems ${\rm C}_{20}{\rm H}_{42}$, ${\rm C}_{100}{\rm H}_{202}$ and ${\rm C}_{1400}{\rm H}_{2802}$.
The solid lines are fits to the numerical data using equation (3) with the parameters $\alpha$ and $T_c$ given in table \ref {table_nT} .
}
\end{figure}

\begin{figure}
  \includegraphics[width=0.3\textwidth, angle=0]{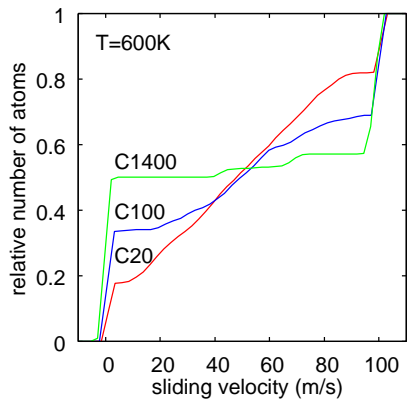}
  \caption{ \label{n_V_d}
(Color online) The relative number of atoms moving at the velocity below $v$, as a function of $v$, at 600 K.
It can be seen that the velocity gradient is going from nearly Newtonian in the case of ${\rm C}_{20}{\rm H}_{42}$ to solid-like
in the ${\rm C}_{1400}{\rm H}_{2802}$ case.
}
\end{figure}

The velocity dependence of the MD data in Fig. \ref{visc_vel_temp} can be very well
fitted by the formula $\eta_{\rm eff} = B \dot \gamma^{-n}$ or
$${\rm log} \eta_{\rm eff} = C - n{\rm log} \dot \gamma,\eqno(2)$$
where $C= {\rm log} B$.
This is illustrated in Fig. \ref{fit} for two cases. The rest of the cases also show that a linear fit is appropriate with the exceptions
of the results at very high temperatures. The ${\rm C}_{100}{\rm H}_{202}$ system
shows the highest effective viscosity due to entanglement. The ${\rm C}_{20}{\rm H}_{42}$ is nearly liquid above 300 K and shows less
entanglement whereas the sliding in ${\rm C}_{1400}{\rm H}_{2802}$ takes place at nearly one interface (see figure \ref{n_V_d}).
We find that the parameters $C$ and $n$ depend on the temperature.
The data points in Fig. \ref{n_T} show the temperature dependence of the index $n$, while the solid
lines are fits to the data points using the inverse sigmoidal curve:
$$n={1\over 1+(T/T_c)^\alpha}.\eqno(3)$$
The parameters $\alpha$ and $T_c$ are given in table \ref {table_nT}.
At low temperatures $n=1$ as expected for dry friction. That is, the frictional shear stress
$\sigma_{\rm f} = \eta_{\rm eff} \dot \gamma = 10^C \dot \gamma^{1-n}$ is independent of the shear rate when
$n=1$ as expected for dry friction at low temperatures (no thermally activated creep). At high temperature
$n$ approaches 0 as expected for a Newtonian fluid where the frictional shear stress is proportional to the shear rate.

\begin{table}
\begin{tabular}{|c|c|c|c|}
  \hline
   & ${\rm C}_{20}{\rm H}_{42}$ & ${\rm C}_{100}{\rm H}_{202}$ & ${\rm C}_{1400}{\rm H}_{2802}$ \\ \hline
  $T_c$ (K) & 353 $\pm$ 28 & 642 $\pm$ 22 & 840 $\pm$ 30 \\ \hline
  $\alpha$ (-) & 4.09 $\pm$ 1.24 & 3.68 $\pm$ 0.46 & 2.79 $\pm$ 0.30 \\
  \hline
  \end{tabular}
\caption{ Table showing the $T_c$ and $\alpha$ in equation (3) for each system.
The standard deviation of the fitting parameters is also indicated. }
\label {table_nT}
\end{table}

\begin{figure}
  \includegraphics[width=0.3\textwidth, angle=0]{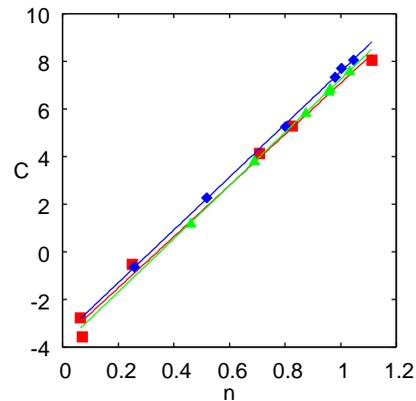}
  \caption{ \label{C_n}
(Color online) The $C$ parameter in equation (2) also follows a sigmoidal curve when the temperature is varied.
This figure shows the dependence of $C$ on $n$. All three systems
${\rm C}_{20}{\rm H}_{42}$, ${\rm C}_{100}{\rm H}_{202}$ and ${\rm C}_{1400}{\rm H}_{2802}$ are represented.
}
\end{figure}

\begin{figure}
  \includegraphics[width=0.3\textwidth, angle=0]{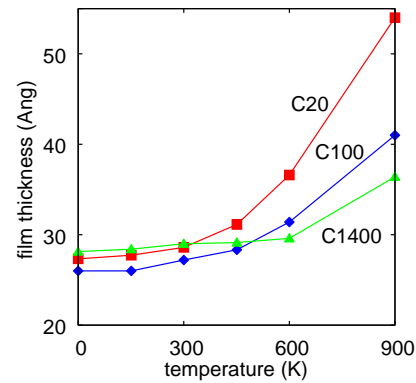}
  \caption{ \label{d_temp}
(Color online) The polymer film thickness as a function of the temperature. The sliding speed is 10 m/s.
All three systems ${\rm C}_{20}{\rm H}_{42}$, ${\rm C}_{100}{\rm H}_{202}$ and ${\rm C}_{1400}{\rm H}_{2802}$ are represented.
}
\end{figure}

The parameter $C$ depends on the units used for $\eta_{\rm eff}$ and $\dot \gamma$ and here we assume SI units.
Remarkably we find a linear relation between $C$ and $n$ for all the systems and temperatures we have studied, see
Fig. \ref{C_n}. The extrapolation of $C$ to $n=0$ gives $C\approx -3.8 \pm 0.2$. This gives the effective viscosity
$\eta_{\rm eff} = 10^C = (1.6-2.5) \times 10^{-4} \ {\rm Pas}$. Experiments have shown \cite{Francis} that for a wide range
of fluids the fluid viscosity at the boiling point is $\approx 2.2 \times 10^{-4} \ {\rm Pas}$, i.e., nearly the same
as we deduce for confined fluids when extrapolating our data to $n=0$ (Newtonian fluid). This result is extremely
interesting but perhaps not entirely unexpected, as $n=0$ corresponds to high temperatures where the separation
between the solid walls (see Fig. \ref{d_temp} in which the sliding velocity is 10 m/s) is much larger than for lower
temperatures and where the mobility of the fluid molecules may be similar to that in the bulk fluid close to the boiling point.
The ${\rm C}_{100}{\rm H}_{202}$ film has a smaller film thickness
than the other systems at low temperatures due to the fact that it only has 6 monolayers at these conditions. The
other two systems have 7 monolayers at the same conditions and have thus thicker films. This matter has been discussed in reference
\cite{SivebaekLangmuir2010}.

Fig. \ref{n_T} shows that the longer the hydrocarbon chain length the higher the temperature necessary for the film
to behave as a Newtonian fluid. This is also illustrated in Fig. \ref{n_V_d}
which shows the relative number of atoms traveling at velocities below $v$, as a function of $v$, for all three polymer films at 600 K.
It can be seen that the velocity gradient is going from nearly Newtonian in the case of ${\rm C}_{20}{\rm H}_{42}$ to solid-like
in the ${\rm C}_{1400}{\rm H}_{2802}$ case.

Note that for the hydrocarbons ${\rm C}_{100}{\rm H}_{202}$ and ${\rm C}_{1400}{\rm H}_{2802}$ at room temperature
$n\approx 0.9$, as also found experimentally for many confined fluids. When $n\approx 0.9$ from Fig. \ref{C_n} we get
$C\approx 6$ which is a little larger than found experimentally (at room temperature) $C_{\rm exp} \approx 5$.
This may reflect somewhat different confinement condition, e.g., differences in contact pressures
($10 \ {\rm MPa}$ in our study as compared to
a few MPa in most of the experimental studies). However, in both cases the contact pressure is so small that one does not expect any
significant dependency of the shear stress on the contact pressure \cite{SivebaekEPJE2008}, except
if there is a pressure induced change in the number of confined layers.

It is well known that the viscosity of fluids at high pressures may be many orders of magnitude larger than at low pressures.
Using the theory of activated processes, and assuming that a local molecular rearrangement in a fluid results in a local volume expansion,
one expects an exponential dependence of viscosity $\eta$ on the hydrostatic pressure $p$, $\eta = \eta_0 {\rm exp}(p/p_0)$, where typically
(for hydrocarbons or polymer fluids) $p_0 \approx 10^8 \ {\rm Pa}$ (see, e.g., Refs. \cite{Eyring,BookP}). Here we are interested in
(wetting) fluids confined between the surfaces of elastically soft solids, e.g., rubber. In this case the pressure at the interface
is usually at most of the order of the Young's modulus, which (for rubber) is less than $10^7 \ {\rm Pa}$.
Thus, in most cases involving elastically soft materials, the viscosity
can be considered as independent of the local pressure.

One of us have recently studied rubber friction on rough surfaces \cite{BorisP}. For unfilled styrene butadiene rubber we found the transfer of a thin
smear film to the substrate. In this case the shear during sliding may be localized to a thin (a few nanometer)
interfacial layer which may exhibit frictional properties very similar to what we have observed in our simulations.
Indeed the experimental data indicated a frictional shear stress of the form predicted above with $n \approx 0.91$ and
$10^C \approx 1.3\times 10^5$ (in SI units), in close agreement with the result of our simulations.

To summarize, we have presented results of molecular dynamics calculations of friction performed for a
block sliding on a substrate separated by $\approx 3 \ {\rm nm}$ thick polymer films
where the alkanes had 20, 100 and 1400 carbon atoms. In all cases we found that the logarithm of the effective viscosity
is proportional to the logarithm of the shear rate, ${\rm log} \eta_{\rm eff} \approx C- n {\rm log} \dot \gamma$.
The index
$n$ varies from 1 (solid-like friction) at very
low temperatures to 0 (Newtonian liquid) at very high temperatures, following an inverse
sigmoidal curve. The $C$ parameter is proportional to $n$ and as $n\rightarrow 0$, $10^C$ extrapolates to the viscosity of the
bulk fluid at the boiling point. At room temperature the parameters $n$ and $C$ have been found to be close to what has
been observed experimentally for a large number of fluids.


Two of the authors (I.M.S. and V.N.S.) acknowledge support from IFF,
FZ-J\"ulich, hospitality and help of the staff during their research visits. I.M.S. acknowledges the 2010 Jacob
Wallenberg Prize in Materials Science from the Royal Swedish Academy of Engineering Sciences.


\begin{thebibliography}{999}
\bibitem{Temp}
The ``temperature'' was obtained from the total kinetic energy but even at the highest sliding velocity
$100\ {\rm m/s}$ the contribution from the center of mass motion to the ``temperature'' is not very large,
corresponding to an increase in the ``temperature'' by only $\sim 15 \ {\rm K}$,
and much smaller at the lower sliding velocities (the contribution to the ``temperature'' from the translational
motion scales as the square of the sliding velocity).
\end{thebibliography}
\end{document}